\documentclass[twocolumn,aps,prc,showpacs,floatfix]{revtex4}
\usepackage{graphicx}
\usepackage{dcolumn}
\usepackage{bm}

\begin{document}

\title{Pion productions by proton and Helium-3 on $^{197}$Au
target\\
at beam energies of 2.8, 5, 10 and 16.587 GeV/nucleon}
\author{Gao-Chan Yong}
\affiliation{Institute of Modern Physics, Chinese Academy of
Sciences, Lanzhou 730000, China}
\author{Xurong Chen}
\affiliation{Institute of Modern Physics, Chinese Academy of
Sciences, Lanzhou 730000, China}
\author{Hu-Shan Xu}
\affiliation{Institute of Modern Physics, Chinese Academy of
Sciences, Lanzhou 730000, China}
\author{Wei Zuo}
\affiliation{Institute of Modern Physics, Chinese Academy of
Sciences, Lanzhou 730000, China}

\begin{abstract}
Based on a Relativistic Boltzmann-Uehling- Uhlenbeck transport
model, proton and $^{3}$He induced reactions on $^{197}$Au target
at beam energies of 2.8, 5, 10 and 16.587 GeV/nucleon are studied.
It is found that compared with proton induced reactions, $^{3}$He
induced reactions give larger cross sections of pion production,
about 5 times those of the proton induced reactions. And more
importantly, pion production from $^{3}$He induced reaction is
more inclined to low-angle emission. Neutrino production via
positively charged pion is also discussed accordingly.
\end{abstract}

\pacs{24.10.Jv, 25.40.-h, 25.55.-e, 24.10.Lx} \maketitle


There exists considerable interest in the possibility that one
type of neutrino may transform into another type while propagation
\cite{wolf78}. It was also argued that neutrino oscillations are
related to stellar collapse \cite{wolf79} and spontaneous parity
non-conservation \cite{rab80}. Nowadays atmospheric
\cite{Ashie:2004mr,Ambrosio:2003yz,Soudan2:2003}, reactor
\cite{Araki:2004mb}, solar neutrino \cite{Smy:2003jf,Ahmed:2003kj}
and accelerator neutrinos \cite{aliu05} provide compelling
evidences for neutrino mass and oscillation. It is interesting to
note that the precise neutrino oscillation parameters are
determined by the KamLAND \cite{abe08} recently. For more about
neutrino physics, please see Ref. \cite{stru103}. In the
accelerator based neutrino experiments, a key issue toward the
development of a muon collider or neutrino beam based on a muon
storage ring is the design of a target/capture system capable of
capturing a large number of pions. These pions then proceed into a
decay channel where the resultant muon decay products are
harvested before being conducted into a cooling channel and then
subsequently accelerated to the final energy of the facility
\cite{joha00}. Understanding of the production of pions in proton
interactions with nuclear targets is thus essential for
determining the flux of neutrinos in accelerator based neutrino
experiments \cite{chem02,chem08}. A large amount of data were
collected by the HARP Collaboration experiments for the above
physical subjects recently \cite{harp09}. In fact, in accelerator
based neutrino experiments, some times one needs Helium-3 induced
reaction. There are many simulation methods focus on such studies,
such as the phenomenological Monte Carlo generators GEANT4
\cite{geant4} and MARS \cite{mars} and other theoretical works
that considering more physical processes
\cite{gallmeister,bert,bert1,bert2,QGSP,casca,iss}. In this
article, after checking the reliability a relativistic transport
model (ART) we made comparative studies of pion production in
proton and $^{3}$He induced reactions on $^{197}$Au target at
incident beam energies of 2.8, 5, 10 and 16.587 GeV/nucleon. And
finally, we simply discussed neutrino production via positively
charged pion.


The well-known Boltzmann-Uehling-Uhlenbeck (BUU) model
\cite{bertsch} has been very successful in studying heavy-ion
collisions at intermediate energies. The ART (A Relativistic
Transport) model \cite{li95} is the relativistic form of the BUU
model, in order to simulate heavy-ion collisions at higher
energies, some new physics were added. It includes baryon-baryon,
baryon-meson, and meson-meson elastic and inelastic scatterings.
The ART model includes the following baryons
$N,~\Delta(1232),~N^{*}(1440),~N^{*}(1535),~\Lambda,~\Sigma$, and
mesons $\pi,~\rho,~\omega,~\eta,~K$, as well as their explicit
isospin degrees of freedom. Both elastic and inelastic collisions
among most of these particles are included. For baryon-baryon
scatterings, the ART model includes the following inelastic
channels: $NN \leftrightarrow N (\Delta N^*)$, $NN \leftrightarrow
\Delta (\Delta N^*(1440))$, $NN \leftrightarrow NN (\pi \rho
\omega)$, $(N \Delta) \Delta \leftrightarrow N N^*$, and $\Delta
N^*(1440) \leftrightarrow N N^*(1535)$. In the above, $N^*$
denotes either $N^*(1440)$ or $N^*(1535)$, and the symbol $(\Delta
N^*)$ denotes a $\Delta$ or an $N^*$. For meson-baryon
scatterings, the ART model includes the following reaction
channels for the formation and decay of resonances: $\pi N
\leftrightarrow (\Delta N^*(1440)~ N^*(1535))$, and $\eta N
\leftrightarrow N^*(1535)$. There are also elastic scatterings
such as $(\pi \rho) (N \Delta N^*) \rightarrow (\pi \rho) (N
\Delta N^*)$. For meson-meson interactions, the ART model includes
both elastic and inelastic $\pi\pi$ interactions, with the elastic
cross section consisting of $\rho$ meson formation and the
remaining part treated as elastic scattering. Also included are
reaction channels relevant to kaon production. The extended ART
model is one part of AMPT (A Multi-Phase Transport Model) model
\cite{ampt}. We use the Skyrme-type parametrization for the mean
field which reads \cite{li95}
\begin{equation}
U(\rho)=A(\rho/\rho_{0})+B(\rho/\rho_{0})^{\sigma}.
\end{equation}
Where $\sigma = 7/6$, A = -0.356 MeV is attractive and B = 0.303
MeV is repulsive. With these choices, the ground-state
compressibility coefficient of nuclear matter K=201 MeV. More
details of the model can be found in the original Ref.
\cite{li95}.

\begin{figure}[th]
\centering
\includegraphics[height=3.8cm,width=0.4\textwidth]{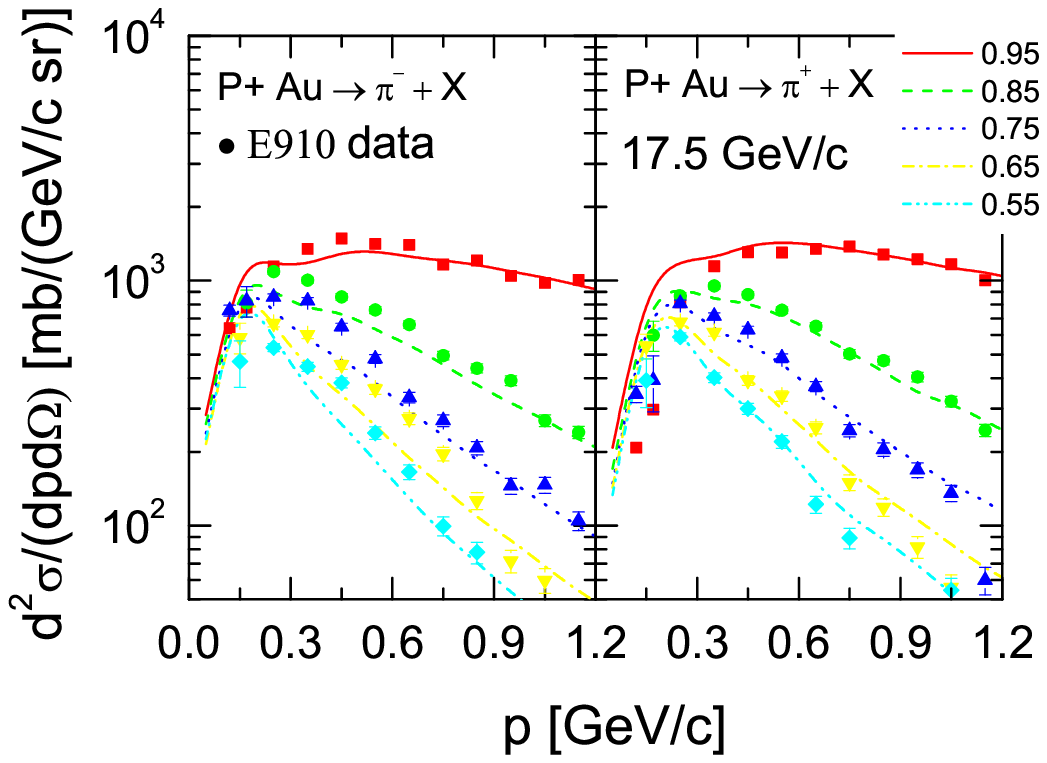}
\includegraphics[height=3.8cm,width=0.4\textwidth]{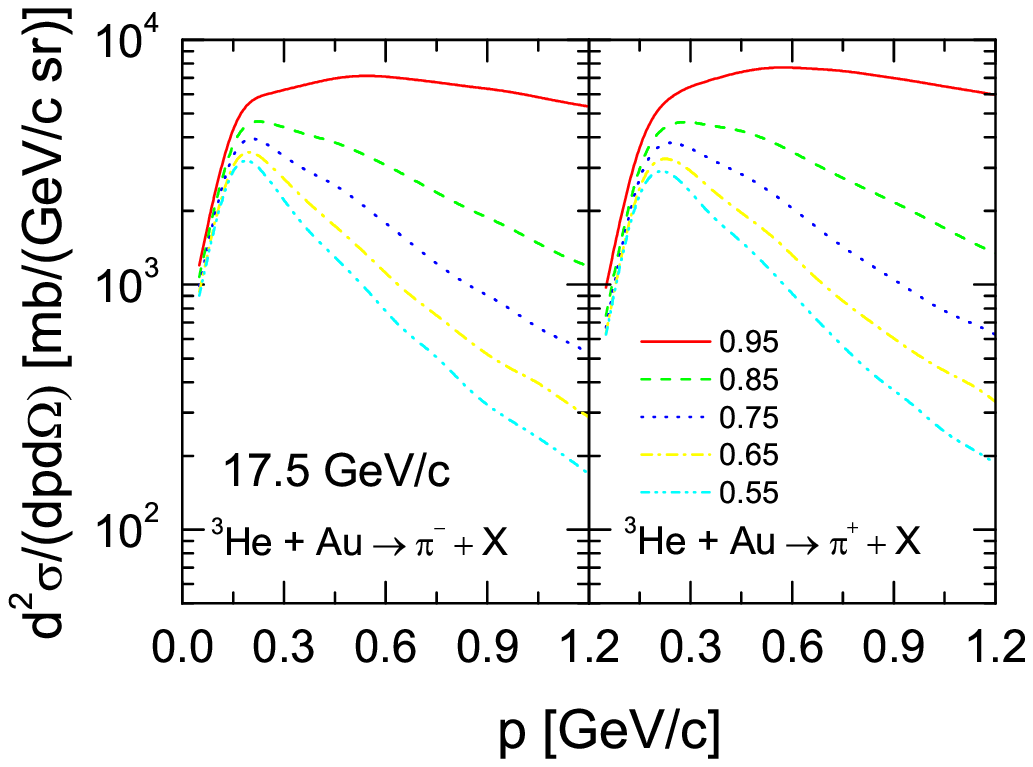}
\caption{(Color online) Top: Production cross sections of
$\pi^{-}$(left column) and $\pi^{+}$(right column) from p+Au at
the incident beam momentum of 17.5 GeV/c ($E_{beam}\sim$ 16.587
GeV/nucleon) shown in bins of $\cos\theta$ (relative to beam
direction). Numbers in the legend refer to the center of each bin.
Data are taken from Ref. \cite{chem02}. Bottom: Same as p+Au case,
but for $^{3}$He+Au.} \label{E910}
\end{figure}
To check the reliability of using the ART model to study the cross
sections of pion production in proton or $^{3}$He induced
reactions, we first made a comparison of pion production in p+Au
reaction at an incident beam momentum of 17.5 GeV/c between the
theoretical simulations and the E910 data \cite{chem02} as shown
in Fig.~\ref{E910}. The top panel of Fig.~\ref{E910} shows the
inclusive differential cross sections of pion production from p+Au
at an incident beam momentum of 17.5 GeV/c. We can see that for
both $\pi^{-}$ and $\pi^{+}$, our results fit the E910 data very
well, especially at higher momenta. Pion production of p+Cu
reaction at incident beam momenta of 12.3 and 17.5 GeV/c also fit
the E910 data \cite{chem02} very well. From Fig.~\ref{E910}, we
can also see that the cross sections at low-angle
($0.9<cos\theta<1$) are evidently larger than those at
high-angles, especially for energetic pion mesons. As a
comparison, we also give the case of $^{3}$He+Au at the incident
beam momentum of 17.5 GeV/c as shown in the bottom panel of
Fig.~\ref{E910}. From the bottom panel of Fig.~\ref{E910}, it is
seen that differential cross sections of pion production of the
$^{3}$He induced reaction are about 5 times those of the proton
induced reaction at the incident beam momentum of 17.5 GeV/c. The
cross sections at low-angle ($0.9<cos\theta<1$) are also much
larger than those at high-angles.


\begin{figure}[t]
\centering
\includegraphics[height=3.8cm,width=0.4\textwidth]{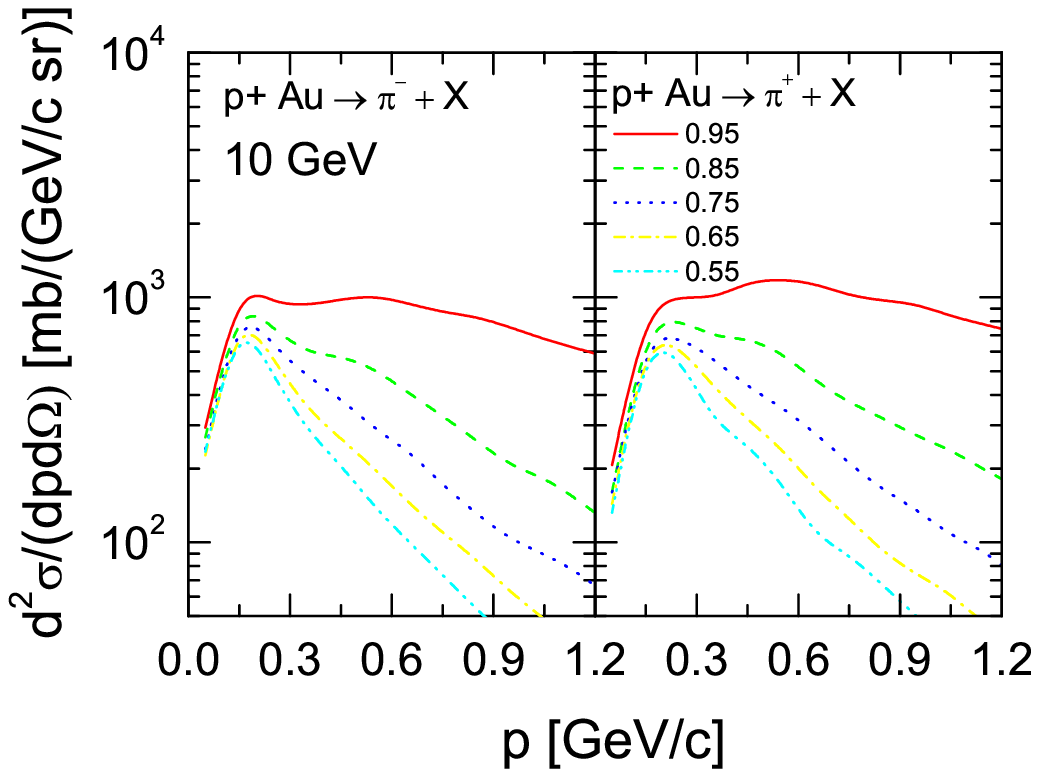}
\includegraphics[height=3.8cm,width=0.4\textwidth]{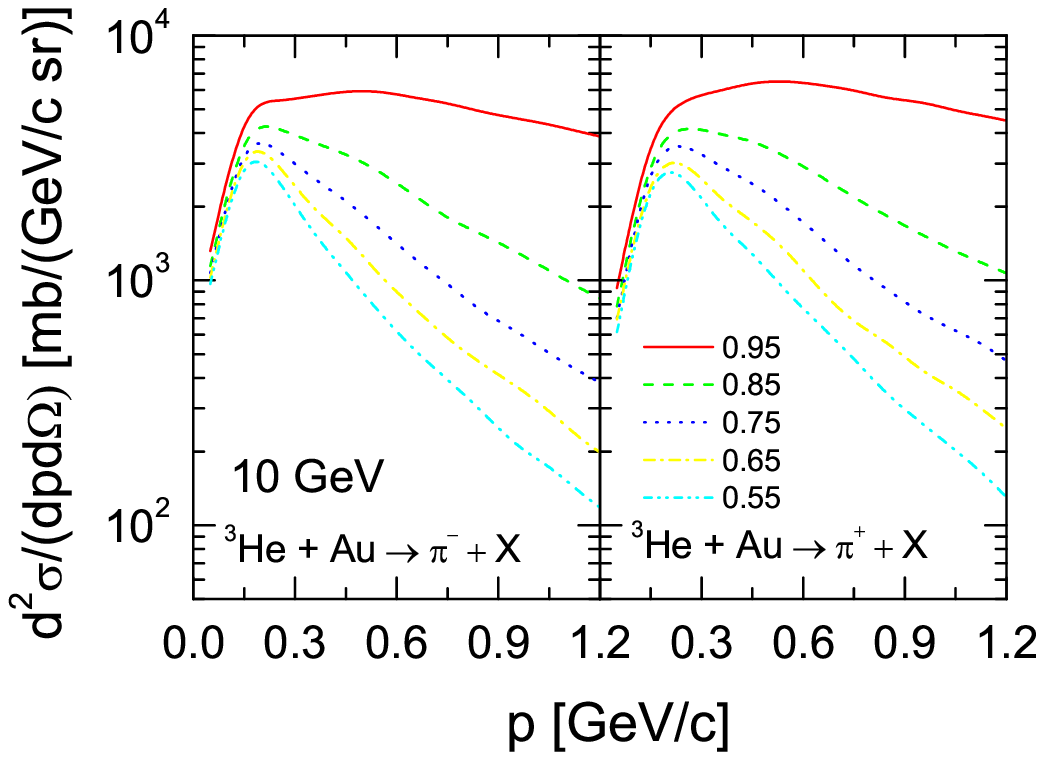}
\caption{(Color online) Top: Production cross sections of
$\pi^{-}$(left column) and $\pi^{+}$(right column) from p+Au at
the incident beam energy of 10 GeV/nucleon shown in bins of
$\cos\theta$. Numbers in the legend refer to the center of each
bin. Bottom: Same as p+Au case, but for $^{3}$He+Au.} \label{E10}
\end{figure}
\begin{figure}[t]
\centering
\includegraphics[height=3.8cm,width=0.4\textwidth]{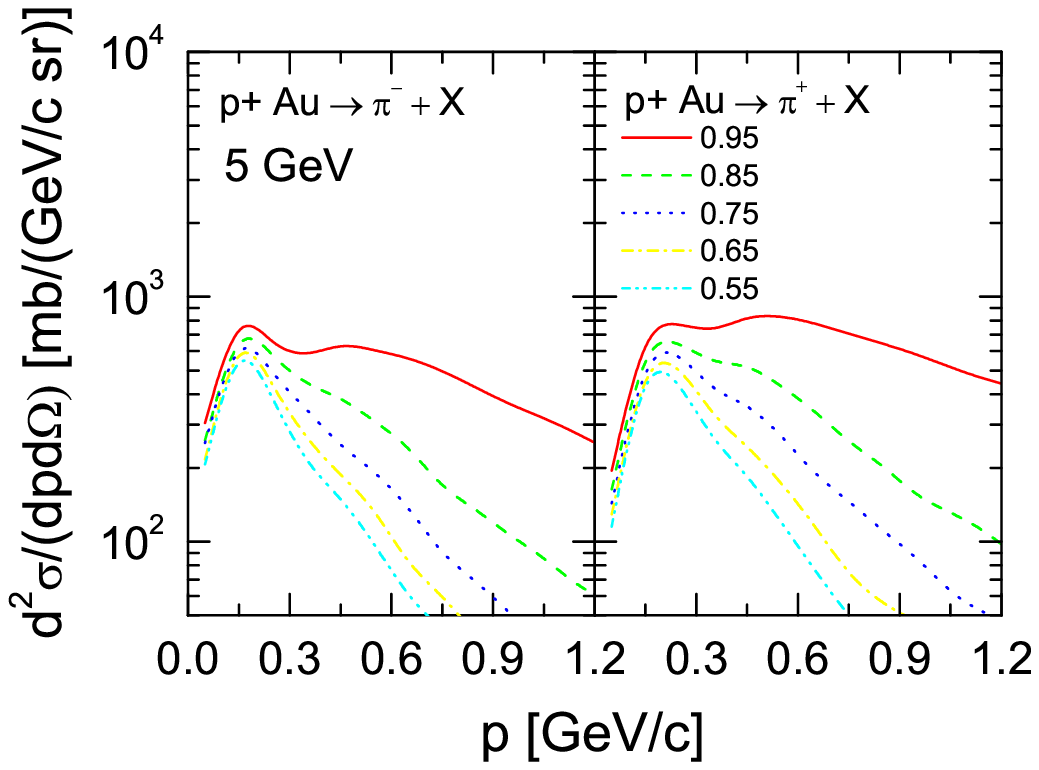}
\includegraphics[height=3.8cm,width=0.4\textwidth]{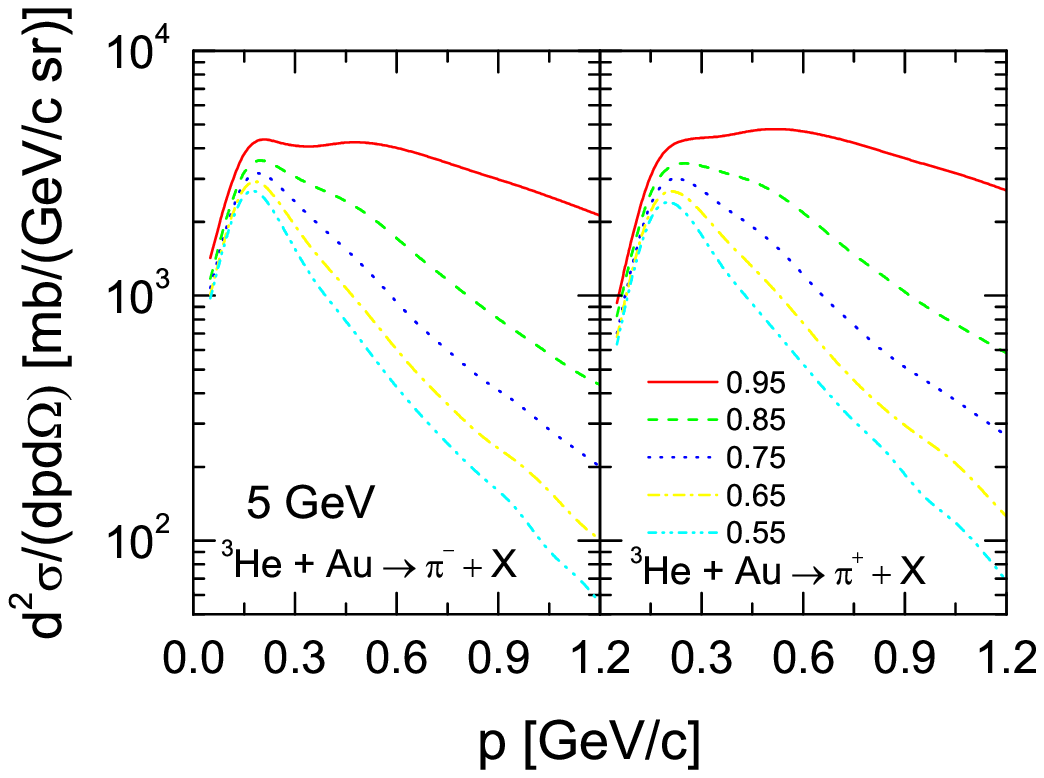}
\caption{(Color online) Top: Production cross sections of
$\pi^{-}$(left column) and $\pi^{+}$(right column) from p+Au at
the incident beam energy of 5 GeV/nucleon shown in bins of
$\cos\theta$. Numbers in the legend refer to the center of each
bin. Bottom: Same as p+Au case, but for $^{3}$He+Au.} \label{E5}
\end{figure}
\begin{figure}[th]
\centering
\includegraphics[height=3.8cm,width=0.4\textwidth]{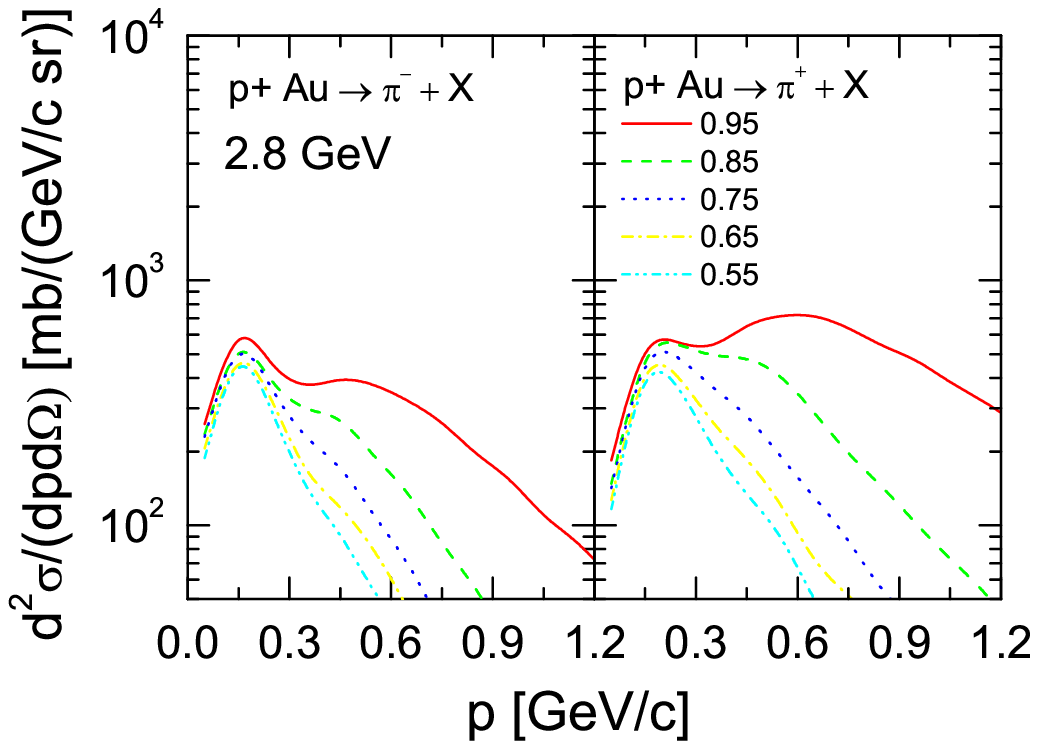}
\includegraphics[height=3.8cm,width=0.4\textwidth]{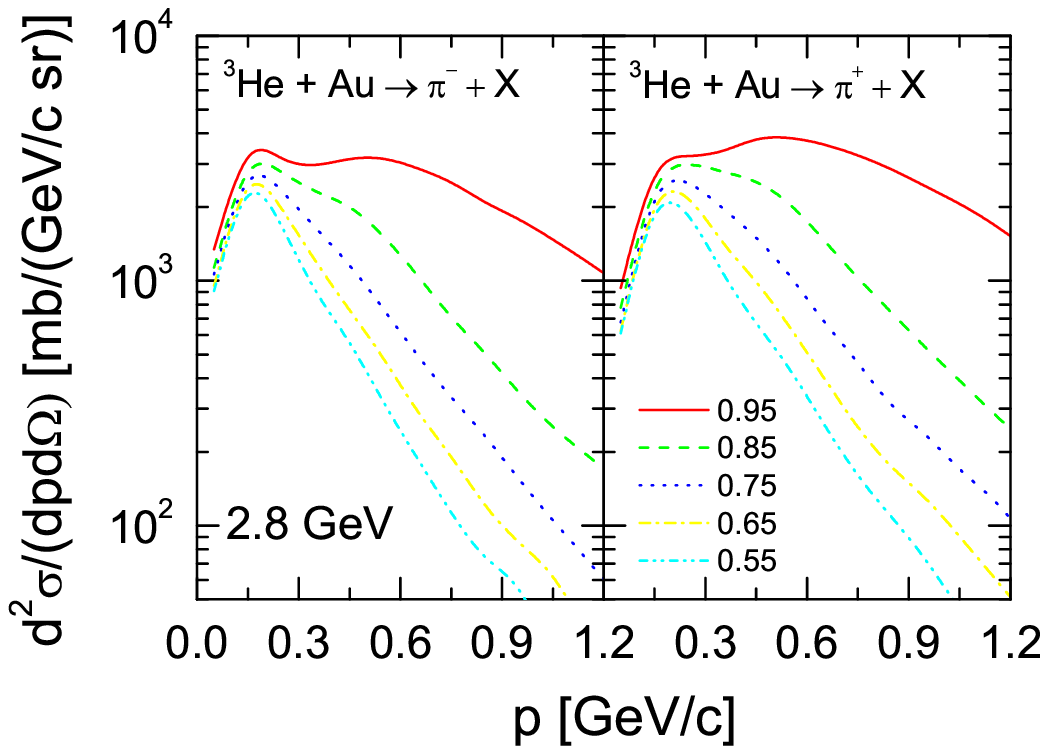}
\caption{(Color online) Top: Production cross sections of
$\pi^{-}$(left column) and $\pi^{+}$(right column) from p+Au at
the incident beam energy of 2.8 GeV/nucleon shown in bins of
$\cos\theta$. Numbers in the legend refer to the center of each
bin. Bottom: Same as p+Au case, but for $^{3}$He+Au.} \label{E2.8}
\end{figure}
To make comparisons systematically between p+Au and $^{3}$He+Au at
different incident beam energies, we plot
Fig.~\ref{E10}-\ref{E2.8}. From the top panels of
Fig.~\ref{E10}-\ref{E2.8}, we can clearly see that as incident
beam energy decreases, cross sections of pion production of p+Au
also decrease. This is understandable since pion production mainly
comes from decays of resonances and energetic nucleon-nucleon
collisions give more resonances. We can also see that as beam
energy decreases, the energetic pion mesons also decrease rapidly.
The bottom panels of Fig.~\ref{E10}-\ref{E2.8} are the cases of
$^{3}$He+Au at different incident beam energies. Also it is seen
that as beam energy decreases, differential cross sections of pion
production decrease rapidly, especially for energetic pion mesons.
Compared with p+Au, as the increase of incident beam energy, cross
sections of pion production at low-angle of $^{3}$He+Au increase
more rapidly, especially for energetic pion mesons. In the
incident beam energy region from 2.8 to 16.587 GeV/nucleon, cross
sections of pion production at low-angle ($0.9<cos\theta<1$) and
high momentum of $^{3}$He+Au are 5$\sim$10 times those of p+Au
case. Using the AMPT model we also made simulations for p+Au at
incident beam energies from 50 to 100 GeV/nucleon, cross sections
of pion production at low-angle ($0.9<cos\theta<1$) are both about
20 times those of p+Au at incident beam energy of 16.587
GeV/nucleon, indicating the saturation of cross section of pion
production at the beam energy of about 50 GeV/nucleon.

\begin{figure}[t]
\begin{minipage}{0.5\textwidth}
\centering
\includegraphics[width=3.99cm]{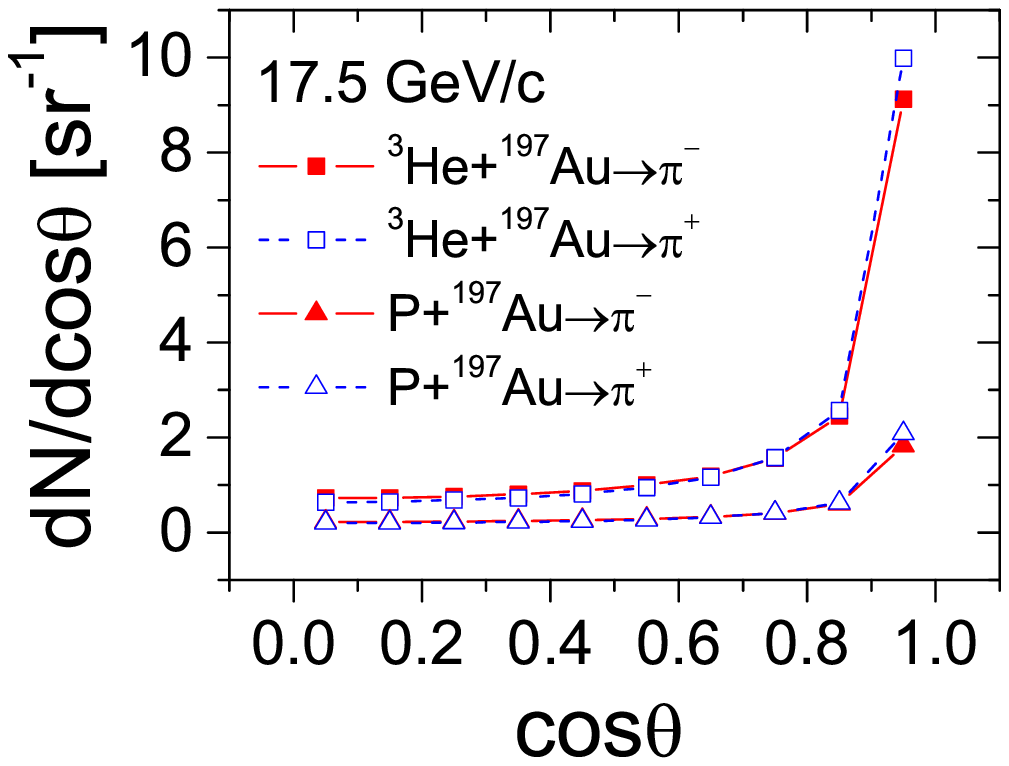}
\includegraphics[width=3.99cm]{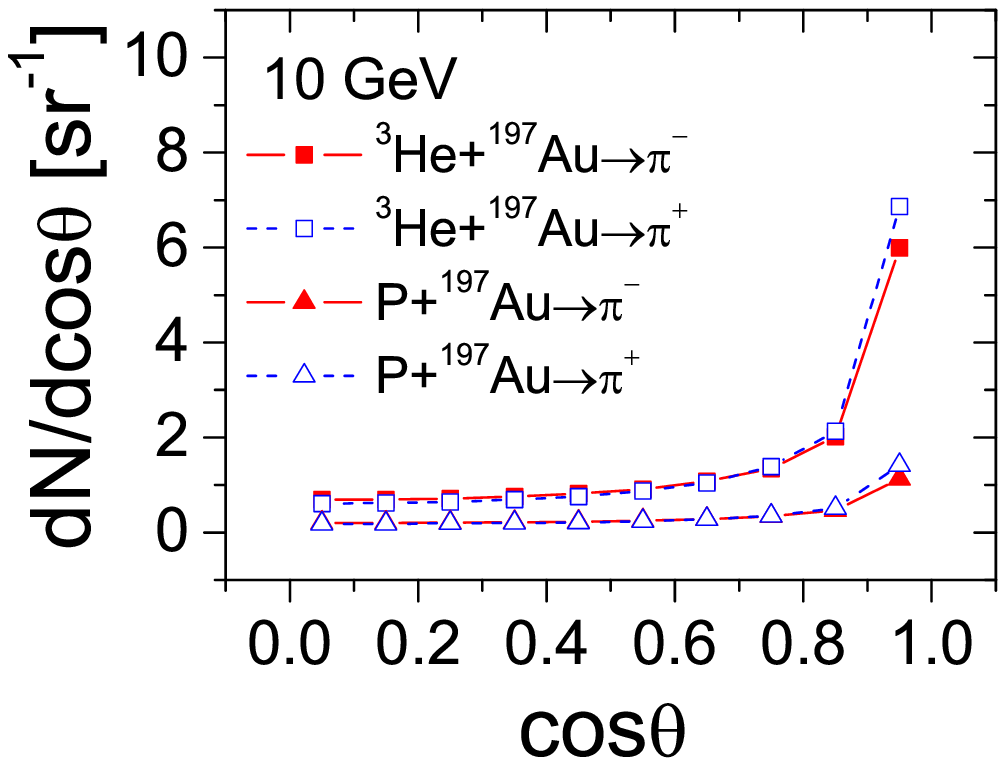}
\end{minipage}
\begin{minipage}{0.5\textwidth}
\includegraphics[width=3.99cm]{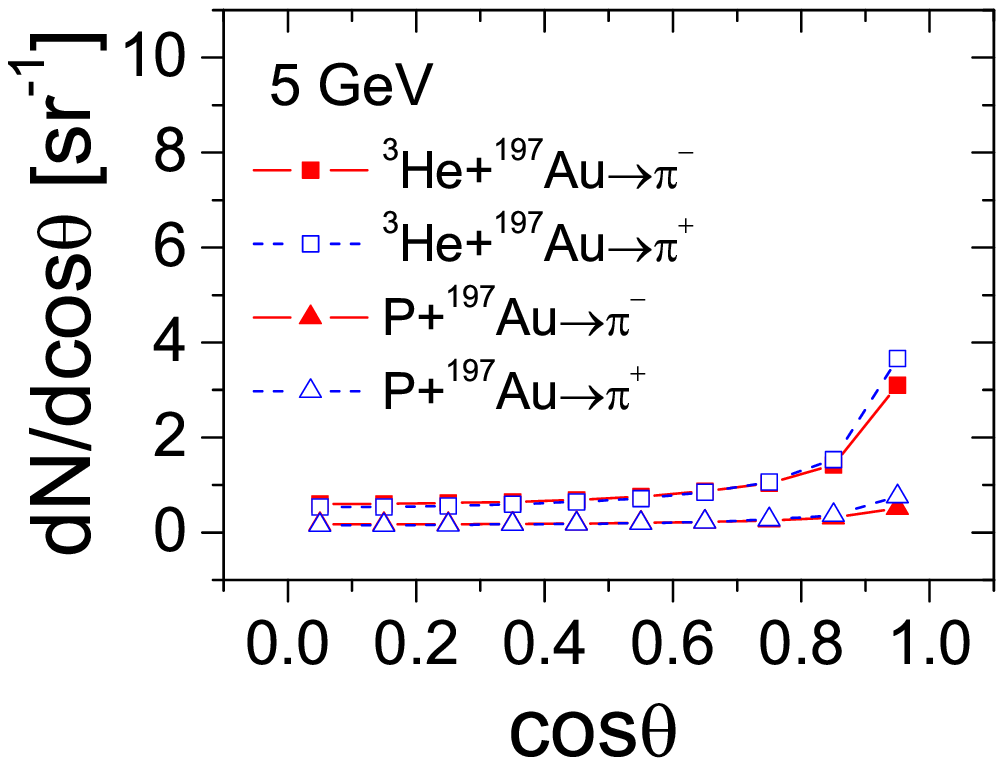}
\includegraphics[width=3.99cm]{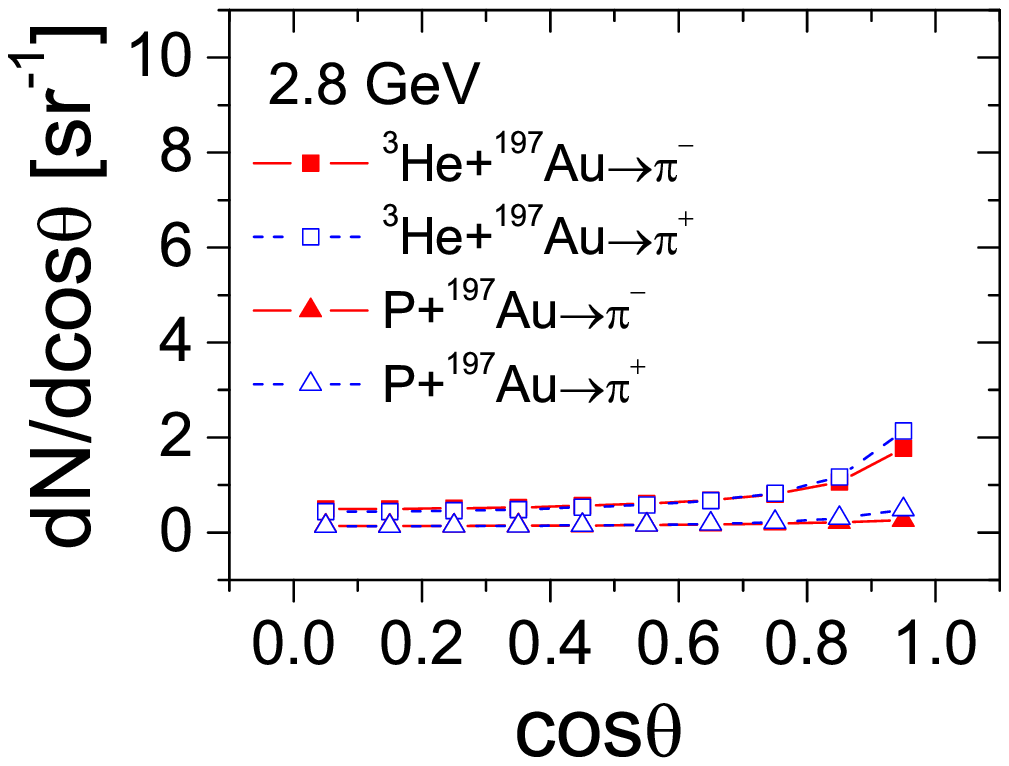}
\caption{(Color online) Angle distributions of pion multiplicity
of p+Au and $^{3}$He+Au at different incident beam energies, in
bins of $\cos\theta$.} \label{cosL}
\end{minipage}
\end{figure}
We next turn to the study of angle distributions of pion
multiplicity of p+Au and $^{3}$He+Au at different incident beam
energies. From Fig.~\ref{cosL}, we can see that whether for p+Au
or $^{3}$He+Au, pion emission at low-angle increases rapidly,
especially at higher incident beam energies. From these plots, we
can also see that the low-angle's pion emission is more pronounced
for $^{3}$He+Au than p+Au, especially at higher incident beam
energies. Pion numbers from $^{3}$He+Au at low-angle
($0.9<cos\theta<1$) are about 5 times those of p+Au. This
indicates that $^{3}$He+Au at high incident beam energy is more
suitable for neutrino experiments compared with p+Au.
\begin{figure}[t]
\begin{center}
\includegraphics[width=0.45\textwidth]{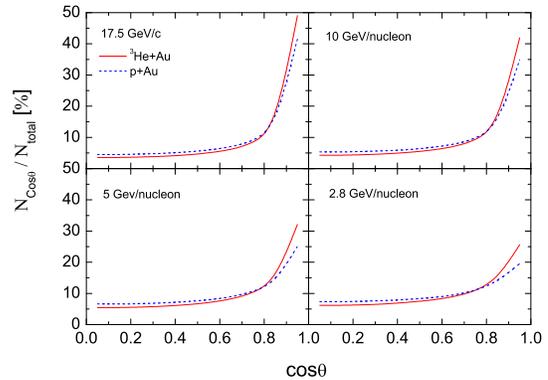}
\end{center}
\caption{(Color online) Angle distributions of charged pion
relative multiplicity ($N_{cos\theta}/N_{total}$) for p+Au and
$^{3}$He+Au reactions at different incident beam energies, in bins
of $\cos\theta$.} \label{relat}
\end{figure}
Fig.~\ref{relat} shows angle distributions of pion relative
emitting number at different incident beam energies. It is seen
that at the high incident beam momentum 17.5 GeV/c, relative
emitting number at low-angle ($0.9<cos\theta<1$) can reach about
50\% for $^{3}$He+Au. While at incident beam energy 2.8
GeV/nucleon, relative emitting number at low-angle
($0.9<cos\theta<1$) reaches only about 25\%. At the studied beam
energy region, we can clearly see that the $^{3}$He induced
reaction on Au target causes larger proportional low-angle pion
emission, especially at higher incident beam energies, about
5\%$\sim$ 10\% larger than that of the proton induced reaction on
Au target.

\begin{figure}[t]
\begin{center}
\includegraphics[width=0.45\textwidth]{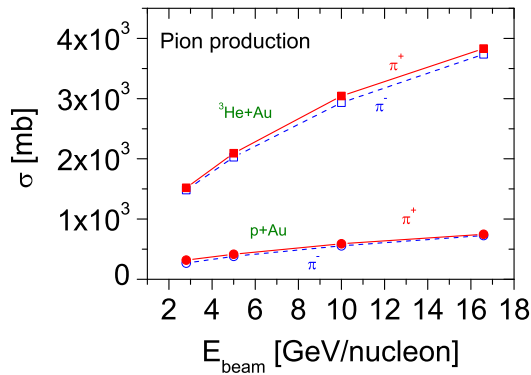}
\end{center}
\caption{(Color online) Cross sections of charged pion production
of p+Au and $^{3}$He+Au reactions at different incident beam
energies.} \label{xsp}
\end{figure}
\begin{figure}[th]
\begin{center}
\includegraphics[width=0.45\textwidth]{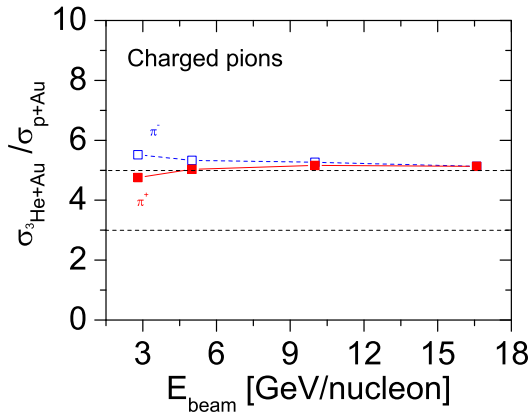}
\end{center}
\caption{(Color online) Ratio of cross sections of charged pion
production of p+Au and $^{3}$He+Au reactions at different incident
beam energies.} \label{xsp1}
\end{figure}
Fig.~\ref{xsp} shows cross sections of charged pion production of
p+Au and $^{3}$He+Au reactions at different incident beam
energies. We can see that at the incident beam energies studied
here, cross sections of pion production of $^{3}$He and
proton-induced reactions on target Au increase about 3 times. The
cross sections of $^{3}$He-induced reaction on target Au at the
incident beam energy of 2.8 GeV/nucleon are larger than those of
p+Au reaction at the beam energy of 16.587 GeV/nucleon.
Fig.~\ref{xsp1} shows the ratio of cross sections of charged pion
production of p+Au and $^{3}$He+Au reactions at different incident
beam energies. We can clearly see that cross sections of charged
pion production from $^{3}$He induced reaction are about 5 times
(larger than $A_{^{3}He}/A_{H}$= 3) those of the proton induced
reaction.

\begin{figure}[th]
\begin{center}
\includegraphics[width=0.45\textwidth]{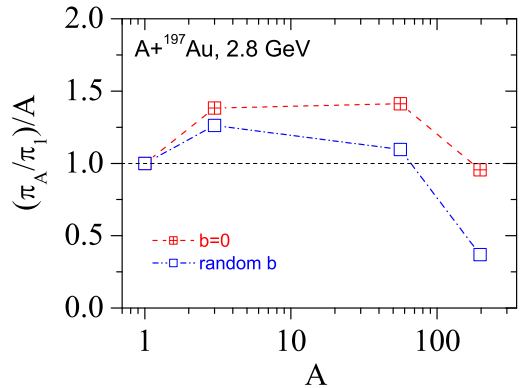}
\end{center}
\caption{(Color online) Ratio of mean pion production per nucleon
of projectile with mass number A and 1 at the incident beam energy
of 2.8 GeV/nucleon. The target is $^{197}$Au and the impact
parameters are set to be 0 and random, respectively.}
\label{scaling}
\end{figure}
Since the work focuses on the proton versus $^{3}$He results, one
wonders what scaling behavior with projectile nucleon number would
one expect from a ``standard'' cascade model (without mean-field
modifications)? Fig.~\ref{scaling} shows the ratio of mean pion
production per nucleon of projectile with mass number A and 1
(proton) at the incident beam energy of 2.8 GeV/nucleon (by the
ART cascade model). We can see that pion production per projectile
nucleon is roughly the same with different projectile mass number
A, i.e., the produced total pion number is roughly proportional to
projectile mass number A. This indicates each nucleon in the
projectile excites pion production almost dependently. But for the
projectile which mass number is smaller than the target mass
number, the scaling behavior that total pion number is roughly
proportional to projectile mass number A is not strictly correct.
In fact, the ratio of $\frac{\pi_{A}}{\pi_{1}}/A$ is always larger
than 1, as shown in Fig.~\ref{scaling}. This is because each
nucleon in the projectile does not excite pion production
independently. One nucleon in the projectile may give energy to
the nucleon in the target, but does not produce pion. The other
nucleon in projectile may also collide with the nucleon which has
obtained energy from the other nucleon of the projectile. Thus the
probability of pion production accordingly increases for the other
induced nucleon in the projectile. This correlation of different
incident nucleons of the projectile does not increase linearly
with projectile's mass number due to the marginal collision of the
induced nucleon. Thus we see about 5 times pion production of the
$^{3}$He induced reaction compared with the proton induced
reaction. For random impact parameter case, smaller
$\frac{\pi_{A}}{\pi_{1}}/A$ is due to the marginal collisions of
the induced nucleon in the projectile.

\begin{figure}[th]
\begin{center}
\includegraphics[width=0.45\textwidth]{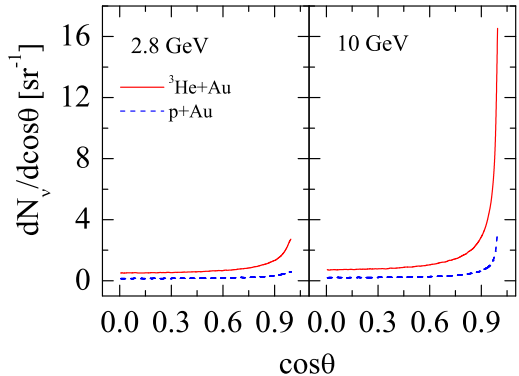}
\end{center}
\caption{(Color online) Angle distributions of neutrino production
from positively charged pion decay in p+Au and $^{3}$He+Au
reactions at incident beam energies of 2.8 and 10 GeV/nucleon, in
bins of $\cos\theta$.} \label{ncd}
\end{figure}
\begin{figure}[t!]
\begin{center}
\includegraphics[width=0.45\textwidth]{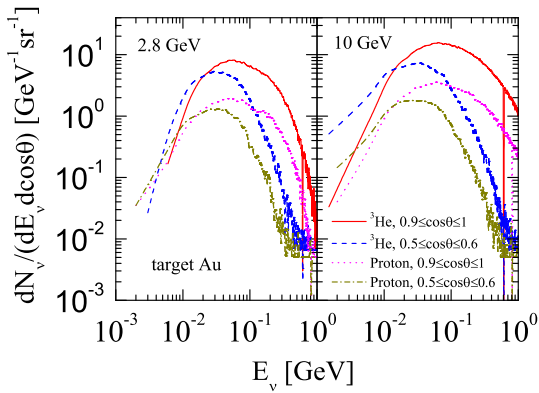}
\end{center}
\caption{(Color online) Energy distributions of neutrino
production at angles $0.9<cos\theta<1$ and $0.5<cos\theta<0.6$
from positively charged pion decay in p+Au and $^{3}$He+Au
reactions at incident beam energies of 2.8 and 10 GeV/nucleon.}
\label{npd}
\end{figure}
To demonstrate neutrino production by proton or $^{3}$He induced
reactions, we plot Fig.~\ref{ncd}, angle distributions of neutrino
via $\pi^{+}$ decay in proton and $^{3}$He induced reactions on
target Au. Assuming $\pi^{+} \rightarrow \mu^{+}+\nu_{\mu}$ and
pion decays into $\mu$ and $\nu_{\mu}$ isotropically in its frame
of reference, we can thus obtain neutrino distribution by assuming
it rest mass 1 eV. Fig.~\ref{ncd} shows angle distributions of
neutrino production from positively charged pion dacay in p+Au and
$^{3}$He+Au reactions at incident beam energies of 2.8 and 10
GeV/nucleon. We can clearly see that neutrinos from $^{3}$He
induced reaction are more inclined to low-angle emission than
proton induced reaction, this situation is clearer for high
incident beam energy. Because the energy distribution of the
emitting neutrinos are important for neutrino-nucleus experiments
\cite{kub94,ek03,lei10,yu11}, we also plot the energy
distributions of the produced neutrinos at low and high angles as
shown in Fig.~\ref{npd}. We can see that the produced neutrinos
possess different energies from about 1 MeV to 1000 MeV and more.
The most probable energy is about 30$\sim$70 MeV for several GeV
incident beam energy. Moreover, we can see that neutrinos from
low-angle possess more energy than those from high-angles. Note
here that neutrino production can also from other channels
\cite{ag09}, especially for energetic collisions. For physical
experiments relevant to neutrinos, detailed studies of the
numbers, the energy spectra as well as the species of emitted
neutrinos are very necessary and therefore the simulations related
to neutrino production are also become important \cite{ag09}.


In conclusion, proton and $^{3}$He induced reactions on $^{197}$Au
target at beam energies of 2.8, 5, 10 and 16.587 GeV/nucleon are
studied in the framework of the Relativistic BUU transport model.
It is found that compared with proton induced reactions, $^{3}$He
induced reactions give larger cross sections of pion production,
about 5 times those of the proton induced reactions. And $^{3}$He
induced reactions are more inclined to low-angle's pion emission.
Simulations demonstrate that neutrino emission via positively
charged pion decay is also inclined to low-angle emission.

The author G.C. Yong acknowledges B.A. Li for providing the ART
model code and Z.W. Lin for providing the new version of the AMPT
model. The work is supported by the National Natural Science
Foundation of China (10635080, 10875151, 10740420550, 10925526),
the Knowledge Innovation Project (KJCX2-EW-N01) of Chinese Academy
of Sciences and the one Hundred Person Project (Y101020BR0), the
Major State Basic Research Developing Program of China under No.
2007CB815004, and the CAS/SAFEA International Partnership Program
for Creative Research Teams (CXTD-J2005-1).

\end{document}